\documentclass[preprint,12pt,sort&compress]{elsarticle}

\usepackage{graphicx}
\usepackage{amsmath}

\begin{document}

\begin{frontmatter}

\title{Effects of final state interactions on charge separation in relativistic heavy ion collisions}

\author{Guo-Liang Ma$^{a}$} \ead{glma@sinap.ac.cn.} 
\author{Bin Zhang$^{b}$} \ead{bzhang@astate.edu.}
            
\address{$^a$ Shanghai Institute of Applied Physics, Chinese Academy of
Sciences, Shanghai 201800, China\\
$^b$ Department of Chemistry and Physics, Arkansas State University, 
P.O. Box 419, State University, AR 72467-0419, USA}

\date{\today}

\begin{abstract}

Charge separation is an important consequence of the Chiral
Magnetic Effect. Within the framework of a multi-phase 
transport model, the effects of final state interactions
on initial charge separation are studied. We demonstrate
that charge separation can be significantly reduced
by the evolution of the Quark-Gluon Plasma produced
in relativistic heavy ion collisions. Hadronization and
resonance decay can also affect charge separation.
Moreover, our results show that the Chiral Magnetic 
Effect leads to the modification of the relation between
the charge azimuthal correlation and the elliptic flow
that is expected from transverse momentum conservation
only. The transverse momentum and pseudorapidity
dependences of, and the effects of background on the 
charge azimuthal correlation are also discussed.

\vspace{1pc}
\end{abstract}

\begin{keyword}
Charge separation  \sep parton
cascade \sep resonance decay 

\PACS 25.75.-q, 24.10.Lx, 25.75.Gz, 25.75.Ld

\end{keyword}

\end{frontmatter}

\section{Introduction}
\label{sec:intro}

Charge separation along the angular momentum direction has been
investigated in relativistic heavy ion collisions 
\cite{:2009uh,:2009txa}.
The experimental study was motivated by the theoretical 
investigation of the Chiral Magnetic Effect 
\cite{Kharzeev:1998kz,Kharzeev:2004ey,
Kharzeev:2007jp,Fukushima:2008xe,Bzdak:2009fc,Fukushima:2010vw,
Basar:2010zd,Muller:2010jd}. 
The Chiral 
Magnetic Effect is related to the fact that the hot and 
dense matter created in heavy ion collisions can form 
${\cal P}$ and ${\cal CP}$ odd metastable domains where 
the parity and time-reversal symmetries are locally violated. 
In the early stage of a non-central relativistic heavy ion 
collision, the magnetic field can reach a magnitude on the 
order of $10^{15}$ T. In the presence of a strong magnetic 
field, these topologically non-trivial domains impose 
constraints on quark chiralities and induce a separation 
of negative and positive particles in the direction of 
magnetic field (i.e. system angular momentum). In spite
of large theory uncertainties, the experimental results
are consistent with Chiral Magnetic Effect expectations.

In addition to the Chiral Magnetic Effect, other effects
can also contribute to charge separation and/or charge
correlation. Bzdak et al. found that the contribution 
due to transverse momentum conservation is comparable in 
magnitude to the prediction of the Chiral Magnetic Effect 
as well as the  data~\cite{Bzdak:2010fd}. Wang also
argued that the measured data can be accounted 
for by cluster particle correlations and new 
physics may not be required to explain 
the data~\cite{Wang:2009kd}. Schlichting 
and Pratt argued that local charge conservation, when 
combined with elliptic flow, explains much of experimental 
measurements~\cite{Schlichting:2010qia}. 
To our knowledge, no previous studies
have included the dynamical effects of final state interactions, 
such as parton cascade and resonance decay, on the experimental 
charge separation observable. On the other hand, these final 
state interaction effects have been found important for many 
experimental observables, such as elliptic flow and particle 
yields. In the following, we will address the problem of
whether an initial charge separation will be able to survive 
the final state interactions.

The paper is organized as follows. In Section \ref{sec:model}, 
we give a brief description of A Multi-Phase Transport (AMPT) 
model and its adaptation for the study of charge separation. 
Results on the charge separation observable are presented 
in Section \ref{sec:resul} followed by a summary in Section 
\ref{sec:concl}.

\section{A Multi-Phase Transport model}
\label{sec:model}

The AMPT model~\cite{Zhang:1999bd,Lin:2001yd,Lin:2004en} 
is a dynamical transport model that 
includes four different stages in relativistic heavy ion 
collisions: the initial condition, partonic interactions, 
the conversion from partonic matter into hadronic matter,
and hadronic rescatterings. The initial condition, which includes 
the spatial and momentum distributions of minijet partons
and soft string excitations, is obtained from the Heavy 
Ion Jet INteraction Generator (HIJING) 
model~\cite{Wang:1991hta,Gyulassy:1994ew}. 
There are two options for doing
the parton evolution and hadronization in the AMPT model.
One option is the default model which includes only interactions
of minijet gluons via Zhang's Parton Cascade 
(ZPC)~\cite{Zhang:1997ej} and uses the Lund string 
fragmentation model~\cite{Sjostrand:1993yb} to turn
partons into hadrons. The other option is the string 
melting model. It starts
the parton evolution with a quark-anti-quark plasma from 
the dissociation of strings. It recombines partons via 
a simple coalescence model to produce 
hadrons~\cite{Lin:2001zk}. 
Dynamics of the subsequent hadronic matter is then 
described by A Relativistic Transport (ART) 
model~\cite{Li:1995pra}.
The default model has good agreement of particle
spectra with experimental data but it significantly 
underestimates the elliptic flow. In contrast, the string 
melting model can only describe low transverse 
momentum spectra, but the agreement with the elliptic
flow data is much better. In addition, the AMPT model
has been used to study other observables, such as
strangeness~\cite{Chen:2006vc,Chen:2006ub}, 
charm~\cite{Zhang:2005ni}, 
J/$\Psi$ production~\cite{Zhang:2000nc,Zhang:2002ug,
Zhang:2006yf}, 
two-pion correlation function~\cite{Lin:2002gc}, 
dijet correlations~\cite{Zhang:2007qx,Pal:2009zz}, 
triangular and higher order flows~\cite{Xu:2010du,Ma:2010dv}.

The AMPT model with string melting starts with 
a quark-anti-quark plasma and it will be used for the 
study of the effects of final state interactions
on charge separation. The Chiral Magnetic Effect 
is not built into the AMPT model. In order to 
separate a fraction of the charges initially, 
we switch the $p_{y}$ values of a fraction of the 
downward moving $u$ quarks with those of the 
upward moving $\bar{u}$ quarks, and likewise for 
$\bar{d}$ and $d$ quarks. The
coordinate system is set up so that the $x$-axis is
in the reaction plane and the $y$-axis is perpendicular
to the reaction plane with the $z$-axis being the
incoming direction of one nucleus. The above
procedure ensures total momentum conservation
while giving momentum kicks to produce an upward
initial current. The current implementation of
the ART model does not conserve the electric charge.
In the following, we turn off the hadron evolution.
The strong parton cascade provides
the most contribution to the evolution and
the exclusion of hadron evolution in the string
melting model is not expected to make significant
changes to the final results. Resonance decays
are implemented to ensure charge conservation
and are included for the study of charge correlations.

\section{Charge separation in heavy ion collisions}
\label{sec:resul}

To measure charge separation possibly coming from local strong 
parity violation in relativistic heavy-ion collisions,
the STAR experiment studied a charge azimuthal correlation observable 
$\left\langle\cos(\phi_{\alpha}+\phi_{\beta}-
2\Psi_{RP})\right\rangle$ as proposed by Voloshin~\cite{Voloshin:2004vk}. 
Here, $\alpha$ and $\beta$ represent the signs of electric charges and
can be positive or negative, while $\Psi_{RP}$ is the azimuthal angle
of the reaction plane.
By measuring the correlations for same-charge and 
opposite-charge pairs, the data show some hints of charge separation, 
which is consistent with the expectation of the Chiral Magnetic 
Effect \cite{:2009uh,:2009txa}. 
In the following, we will look at how the initial
charge separation contributes to the charge azimuthal correlation.
The reaction plane azimuthal angle will be set to zero degrees.
Au+Au collisions at $\sqrt{s_{_{\rm NN}}}=200$ GeV with 
a 10 mb parton cross will be studied and light quark
and charged pion correlations will be analyzed.

\begin{figure}
\center{
\includegraphics[scale=0.85]{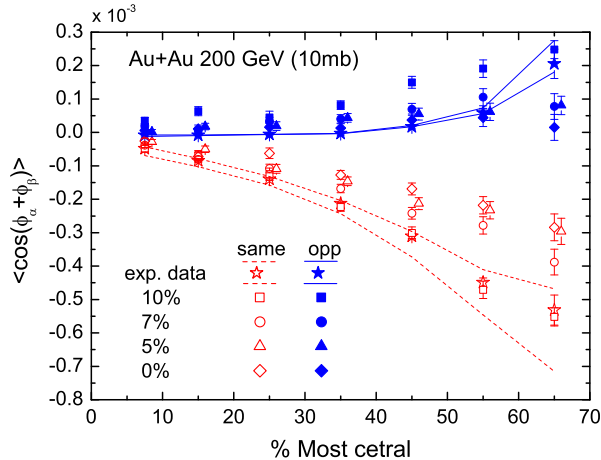}
}
\caption{\footnotesize Centrality dependence of 
$\left\langle\cos(\phi_{\alpha}+\phi_{\beta})\right\rangle$ 
in Au+Au collisions at $\sqrt{s_{_{\rm NN}}}$=200 GeV 
(with a 10 mb parton cross section). 
The different symbols represent different 
percentages of initial charge separation in AMPT calculations. 
The stars represent experimental data, where the two 
surrounding curves give the systematic uncertainty for data. 
Some points are slightly shifted for clarity.
}
\label{fig:cent_dep}
\end{figure}

Fig.~\ref{fig:cent_dep} presents the charge azimuthal 
correlation as a function of centrality from the AMPT simulations.
Since the initial charge separation could depend on centrality, 
different percentages of initial charge separation are
used for each centrality bin to look for possible centrality dependence.
The STAR correlation data are also shown for comparison. 
For the same-charge correlation, results from the AMPT model 
without initial charge separation have smaller magnitudes than data. 
As the percentage of initial charge separation increases,
the magnitude of the correlation increases. The increase is
not linear in the initial charge separation percentage.
A percentage of 10\% for initial charge separation can describe well 
the STAR measurements. For the opposite-charge correlation, 
it seems that initial 
charge separation is not necessary for all centralities except 
the most peripheral bin of 60-70\%. For the centrality bin of 
60-70\%, a percentage of 10\% is indeed needed to match the 
experimental observation. In other words, the observed 
opposite-charge correlation changes much faster than the AMPT results.
Even though 10\% initial charge separation can describe
both the same-charge and opposite-charge correlations for the
60\%-70\% centrality bin, it is difficult to describe the 
centrality dependence of both the same-sign and 
opposite-sign correlations with initial charge separation alone.
We also notice that the results with 5\% initial charge
separation are almost identical to those without initial
charge separation. This indicates that it might be 
challenging to observe an initial charge separation 
of 5\% or less in the presense of strong final state 
interactions.

\begin{figure}
\center{
\includegraphics[scale=0.85]{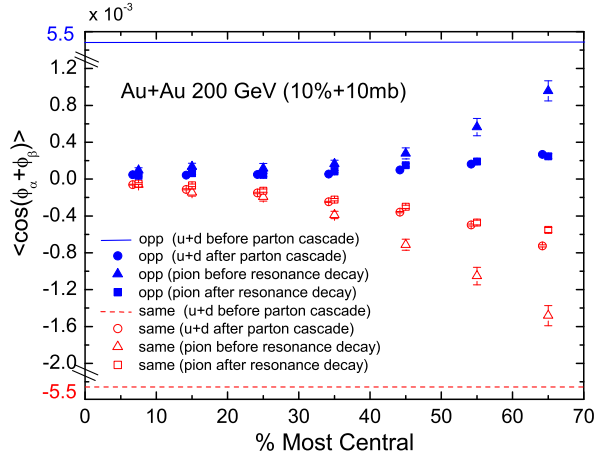}
}
\caption{\footnotesize Centrality dependence of 
$\left\langle\cos(\phi_{\alpha}+\phi_{\beta})\right\rangle$ 
for different stages in AMPT calculations with an initial charge 
separation percentage of 10\% for Au+Au collisions at 
$\sqrt{s_{_{\rm NN}}}$=200 GeV (with a 10 mb
parton cross section). Some points are slightly shifted for clarity. }
\label{fig:evol_sepa}
\end{figure}

To understand how the charge correlation observable evolves in 
heavy-ion collisions, Fig.~\ref{fig:evol_sepa} shows the 
centrality dependence 
of charge correlations for different stages in the AMPT model. 
With an initial percentage of 10\% charge separation, the initial 
charge correlations are quite large (solid and dash lines), with
a magnitude of about $5.5\times10^{-3}$. After strong parton 
cascade, charge correlations are significantly reduced especially 
for central collisions because of frequent parton interactions 
under high parton density. The charge correlations are recovered 
partly from hadronization as coalescence reduces the number
of particles while combining quarks into hadrons.
Resonance decays act opposite to coalescence and reduce charge
correlations in the hadronic phase. The final charged pion correlations
have magnitudes comparable with those of final partons. 
Related to the charge correlation is the percentage of
charge separation. Its centrality dependence has the
same qualitative evolution where parton cascade and resonance
decay decrease while coalescence increases the percentage.
From a percentage of charge separation of 10\% in the beginning, 
only 1-2\% percentage remains at the end with more peripheral 
collisions having larger percentages.

\begin{figure}
\center{
\includegraphics[scale=0.85]{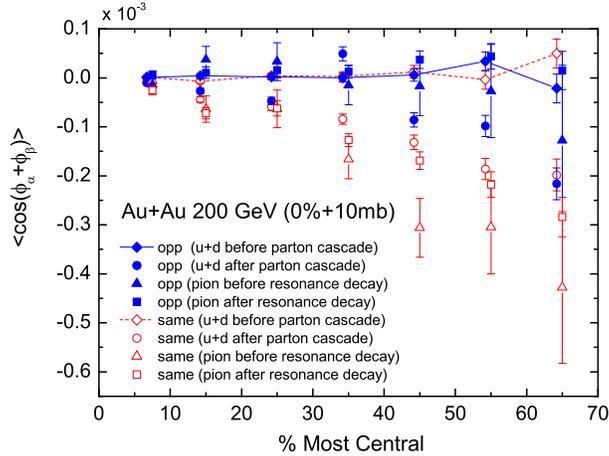}
}
\caption{\footnotesize Centrality dependence of 
$\left\langle\cos(\phi_{\alpha}+\phi_{\beta})\right\rangle$ 
for different stages from AMPT calculations without initial charge 
separation for Au+Au collisions 
at $\sqrt{s_{_{\rm NN}}}$=200 GeV (with a 10 mb parton
cross section). Some points are slightly shifted for clarity.
}
\label{fig:evol_nose}
\end{figure}

As a comparison, the charge correlations at different stages
with no initial charge separation are shown in Fig.~\ref{fig:evol_nose}
for different centrality bins. Before the parton cascade,
both the same-charge and the opposite-charge correlations are 
consistent with zero. After the parton stage, both correlations
become negative with the same-charge correlation having 
the larger magnitude. 
Negative correlations indicate that the correlated charges move
together and they are not separated.
Coalescence increases the magnitude for the same-charge 
correlation and resonance decay decreases it as in
the case with a non-zero initial charge separation. However,
for opposite charges, coalescence reduces the correlation.
If the opposite-charge correlation is calculated including 
charged rho mesons in addition to charged pions, it has a 
magnitude that is larger than that of quarks after 
parton cascade. This shows that when there is no 
initial charge separation, the opposite-charge correlation 
from coalescence is not equally distributed among different 
species combinations.

\begin{figure}
\center{
\includegraphics[scale=0.85]{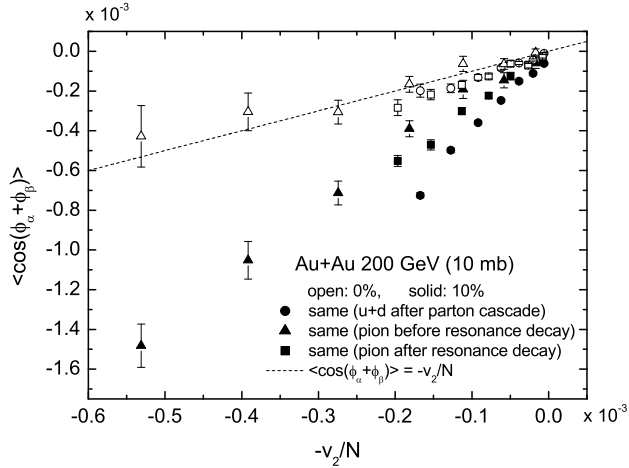}
}
\caption{\footnotesize 
$\left\langle\cos(\phi_{\alpha}+\phi_{\beta})\right\rangle$ as a 
function of $-v_{2}/N$ for different stages in AMPT calculations 
without (0\%, open symbols) and with (10\%, solid symbols) 
initial charge separation for Au+Au collisions at
$\sqrt{s_{_{\rm NN}}}$=200 GeV (with a 10 mb parton cross section). 
The dashed line represents the relation of 
$\left\langle\cos(\phi_{\alpha}+\phi_{\beta})\right\rangle=-v_{2}/N$.
}
\label{fig:evol_v2}
\end{figure}

Recently, Bzdak et al. found that transverse 
momentum conservation can contribute to the charge correlations
with magnitudes comparable to experimentally observed 
correlations~\cite{Bzdak:2010fd}. 
The charge correlations can be calculated
from transverse momentum conservation alone. Under the assumption 
that all particles have the same average transverse momentum, there
is a simple relation between the charge correlations and the
elliptic flow. Both the same-charge and opposite-charge correlations 
are equal to -$v_{2}/N$ for sufficiently large $N$. Here $v_{2}$ 
is the elliptic flow coefficient and $N$ is the total number of 
produced particles (similar results were also obtained 
in \cite{Pratt:2010gy,Pratt:2010zn}). The opposite-charge correlation 
can be affected by factors other than transverse momentum 
conservation. In the following, we will look at how 
this relation holds for the same-charge correlation.
Fig.~\ref{fig:evol_v2} presents the same-charge correlation as 
a function of $-v_{2}/N$ for different stages from AMPT 
calculations without (0\%, open symbols) and with 
(10\%, solid symbols) initial charge separation.
Here $v_{2}$ and $N$ are for particles with pseudorapidity 
$|\eta|<$1. The AMPT results without initial charge 
separation are consistent with the expectation that 
$\left\langle\cos(\phi_{\alpha}+\phi_{\beta})\right\rangle=-v_{2}/N$, 
which is shown in the figure by a dashed line.
It indicates that the same-charge correlation is driven by 
transverse momentum conservation in the AMPT model without 
initial charge separation. On the other hand, the AMPT results 
with initial charge separation are much lower than the expected 
relation. It is interesting to see that the linear relation
between the same-charge correlation and $-v_2/N$ is approximately
preserved with a coefficient much large than $1$. 
Since the AMPT results with 10\% initial charge separation can 
describe the same-charge correlation data well as shown in 
Fig.~\ref{fig:cent_dep}, transverse momentum conservation can only 
partly account for the measured charge correlation data.

\begin{figure}
\center{
\includegraphics[scale=0.85]{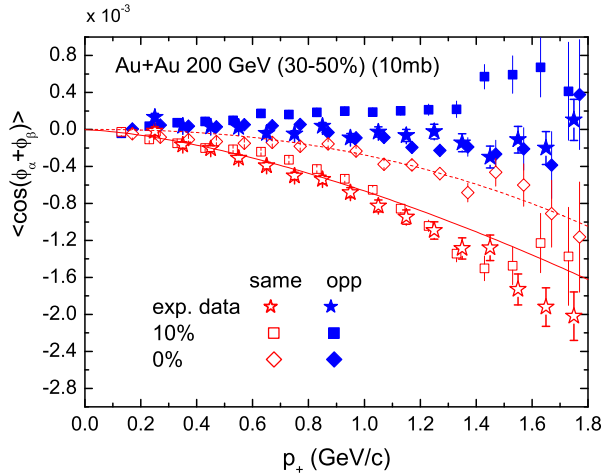}
}
\caption{\footnotesize 
$\left\langle\cos(\phi_{\alpha}+\phi_{\beta})\right\rangle$ as a 
function of $p_{+}$ = ($p_{t, \alpha}+p_{t, \beta}$)/2 in AMPT calculations 
without (0\%, diamonds) and with (10\%, squares) 
initial charge separation for the 30-50\% centrality bin in Au+Au collisions at
$\sqrt{s_{_{\rm NN}}}$=200 GeV (with a 10 mb parton cross section). 
The curves are power-law fits for the same-charge correlations from 
the AMPT calculations without (dash) and with (solid) initial charge 
separation, and the stars represent experimental data. Some points 
are slightly shifted for clarity.
}
\label{fig:evol_pt}
\end{figure}

In more detail, Fig.~\ref{fig:evol_pt} shows the dependences of 
charge correlations on the average of the transverse momentum 
($p_{+}$ = ($p_{t, \alpha}+p_{t, \beta}$)/2) of two final charged 
pions for the 30-50\% centrality bin. For the same-charge 
correlation, the magnitudes of results from the AMPT model 
without initial charge separation are smaller than those of data 
while 10\% initial charge separation can increase the magnitudes 
to reproduce data. For opposite-charge pairs, the correlation 
with no initial charge separation is consistent with data
while the correlation with 10\% initial charge separation 
increases weakly with $p_{+}$ and is a little higher than data.  
All of these results are consistent with the integrated 
correlations which are presented in Fig.~\ref{fig:cent_dep}.  
From transverse momentum conservation,  
$\left\langle\cos(\phi_{\alpha}+\phi_{\beta})\right\rangle$ is 
proportional to $p_{+}^{n}$ with $n=2$ to 3~\cite{Bzdak:2010fd}. 
The curves in Fig.~\ref{fig:evol_pt} show the power-law fits
to the same-charge correlations from the AMPT model. 
The power $n$ is 2.24 $\pm$ 0.27 when there is no 
initial charge separation, consistent with the expectation from 
transverse momentum conservation. However, the power $n$ deceases 
to 1.54 $\pm$ 0.18 when 10\% initial charge separation
is included. In addition, we found that the charge correlations 
depend very weakly on $p_{-}=|p_{t, \alpha}-p_{t, \beta}|$.
In particular, the opposite-charge correlation 
increases gradually to a level of about $0.1\times 10^{-3}$
while the same-charge correlation stays at a constant level of about
$0.25\times 10^{-3}$. Even though the magnitude of the same-charge
correlation is smaller than the experimental data, the integrated
value is consistent with experimental data because the lowest $p_{-}$
bin carries the highest weight.

\begin{figure}
\center{
\includegraphics[scale=0.85]{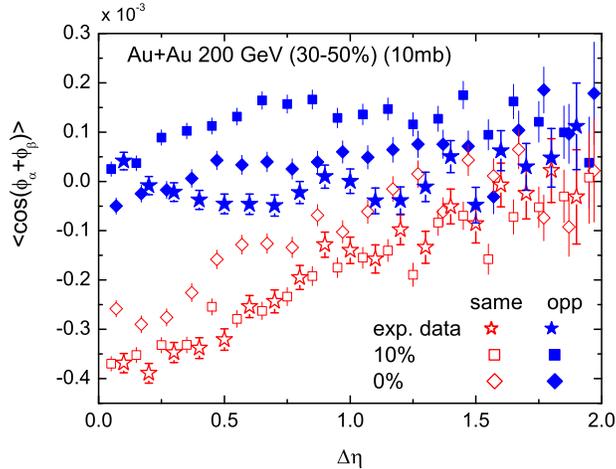}
}
\caption{\footnotesize 
$\left\langle\cos(\phi_{\alpha}+\phi_{\beta})\right\rangle$ as a 
function of $\Delta\eta = |\eta_{\alpha}-\eta_{\beta}|$ in AMPT calculations 
without (0\%, diamonds) and with (10\%, squares) 
initial charge separation for the 30-50\% centrality bin in 
Au+Au collisions at $\sqrt{s_{_{\rm NN}}}$=200 GeV 
(with a 10 mb parton cross section). 
The stars represent experimental data. 
Some points are slightly shifted for clarity.
}
\label{fig:evol_eta}
\end{figure}

Fig.~\ref{fig:evol_eta} presents charge correlations as functions of
the pseudorapidity difference 
($\Delta\eta = |\eta_{\alpha}-\eta_{\beta}|$) 
of two final charged pions for the 30-50\% centrality bin. Again we see
that results from the AMPT model without initial charge separation 
can describe the opposite-charge correlation, while initial charge separation
is needed to reproduce the same-charge correlation data. It is worth 
mentioning that the strong dependence on the pseudorapidity difference 
cannot be obtained in the present calculations from transverse momentum 
conservation~\cite{Bzdak:2010fd}. More realistic longitudinal dynamics
in the AMPT model contributes to the better description of the 
dependence on the pseudorapidity difference. 

\begin{figure}
\center{
\includegraphics[scale=0.85]{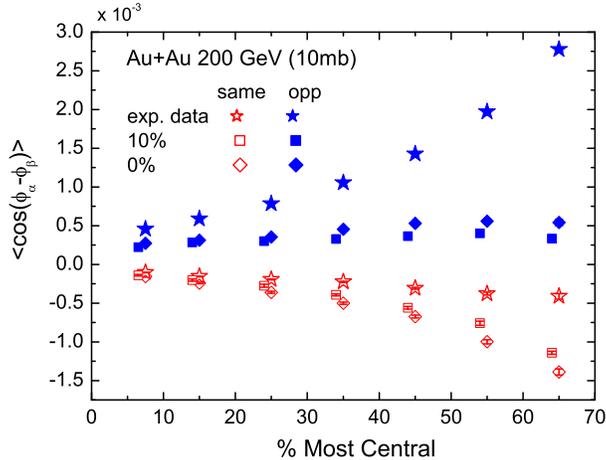}
}
\caption{\footnotesize 
Centrality dependence of 
$\left\langle\cos(\phi_{\alpha}-\phi_{\beta})\right\rangle$ from 
AMPT calculations without (0\%, diamonds) and with (10\%, squares) 
initial charge separation in Au+Au collisions at 
$\sqrt{s_{_{\rm NN}}}$=200 GeV (with a 10 mb parton cross section). 
The stars represent experimental data. 
Some points are slightly shifted for clarity.
}
\label{fig:evol_minus}
\end{figure}

In addition to the angular correlation 
$\left\langle\cos(\phi_{\alpha}+\phi_{\beta})\right\rangle$, 
charge separation also shows up in the angular correlation 
$\left\langle\cos(\phi_{\alpha}-\phi_{\beta})\right\rangle$. 
The former is free of reaction plane independent backgrounds
while the latter is also sensitive to reaction plane independent
backgrounds. Charge separation increases the opposite-charge 
correlation and decreases the same-charge correlation for
the former, while it decreases the opposite-charge correlation
and increases the same-charge correlation for the latter.
We will look at the centrality dependence of the
charge correlation 
$\left\langle\cos(\phi_{\alpha}-\phi_{\beta})\right\rangle$ 
in Fig.~\ref{fig:evol_minus}.
When there is no initial charge separation, the AMPT results 
have the same trends as the experimental
data for both the same-charge and the opposite-charge correlations.
However, the correlations are much lower than
those observed experimentally. Charge separation brings
the same-charge correlation closer to data and the opposite-charge
correlation farther away from data by amounts comparable
to those for 
$\left\langle\cos(\phi_{\alpha}+\phi_{\beta})\right\rangle$.
However, the changes are not enough to make up for the 
large difference between the AMPT results and the experimental
data. Additional backgrounds
that can significantly increase the correlations are needed
in order to describe the data.

\section{Conclusions}
\label{sec:concl}

In summary, final state interactions play an important role on 
charge separation in relativistic heavy-ion collisions. 
Parton cascade and resonance decay significantly reduce the 
charge separation from 10\% in the initial state to 1-2\% 
in the final state. Therefore, it is essential to take these 
final state effects into account for studies related to charge 
separation. Our results also suggest that mechanisms beyond
transverse momentum conservation will be needed even for
the description of the same-charge correlation.

Our approach includes the effects of local charge conservation
and transverse momentum conservation automatically. However,
detailed magnetic field evolution~\cite{Toneev:2010xt}, 
or fluctuating domain
sizes, or different topological charges are not included. 
These effects can lead to different charge separation 
percentages for different centralities. But they are not
likely to help improve the simultaneous description of
both the same-charge and opposite-charge correlations, 
and both $\left\langle\cos(\phi_{\alpha}+\phi_{\beta})\right\rangle$ 
and $\left\langle\cos(\phi_{\alpha}-\phi_{\beta})\right\rangle$. 
Schlichting and Pratt
recently demonstrated that charge balancing can affect
the difference between the opposite-charge and same-charge
correlations~\cite{Schlichting:2010qia}. This and other
possible mechanisms certainly deserve further study for
a satisfactory understanding of experimental data.

\section*{Acknowledgements}
We thank A. Bzdak, C.M. Ko, V. Koch, J. Liao, Y.G. Ma, 
F. Wang for helpful discussions
and the U.S. National Energy Research Scientific 
Computing Center for providing computing resources.
This work was supported by the NSFC of China under Projects 
Nos. 10705044, 11035009, the Knowledge Innovation 
Project of Chinese Academy of Sciences under 
Grant No. KJCX2-EW-N01 (G.L.M.), and by the U.S.
National Science Foundation under Grant Nos. PHYS-0554930 and
PHYS-0970104 (B.Z.).



\begin{thebibliography}{00}

\bibitem{:2009uh}
  B.~I.~Abelev {\it et al.}  [STAR Collaboration],
  Phys.\ Rev.\ Lett.\  {\bf 103} (2009) 251601
  [arXiv:0909.1739 [nucl-ex]].

\bibitem{:2009txa}
  B.~I.~Abelev {\it et al.}  [STAR Collaboration],
  Phys.\ Rev.\  C {\bf 81} (2010) 054908
  [arXiv:0909.1717 [nucl-ex]].

\bibitem{Kharzeev:1998kz}
  D.~Kharzeev, R.~D.~Pisarski and M.~H.~G.~Tytgat,
  Phys.\ Rev.\ Lett.\  {\bf 81} (1998) 512
  [arXiv:hep-ph/9804221].

\bibitem{Kharzeev:2004ey}
  D.~Kharzeev,
  Phys.\ Lett.\  B {\bf 633} (2006) 260
  [arXiv:hep-ph/0406125].

\bibitem{Kharzeev:2007jp}
  D.~E.~Kharzeev, L.~D.~McLerran and H.~J.~Warringa,
  Nucl.\ Phys.\  A {\bf 803} (2008) 227
  [arXiv:0711.0950 [hep-ph]].

\bibitem{Fukushima:2008xe}
  K.~Fukushima, D.~E.~Kharzeev and H.~J.~Warringa,
  Phys.\ Rev.\  D {\bf 78} (2008) 074033
  [arXiv:0808.3382 [hep-ph]].

\bibitem{Bzdak:2009fc}
  A.~Bzdak, V.~Koch and J.~Liao,
  Phys.\ Rev.\  C {\bf 81} (2010) 031901
  [arXiv:0912.5050 [nucl-th]].

\bibitem{Fukushima:2010vw}
  K.~Fukushima, D.~E.~Kharzeev and H.~J.~Warringa,
  Phys.\ Rev.\ Lett.\  {\bf 104} (2010) 212001
  [arXiv:1002.2495 [hep-ph]].

\bibitem{Basar:2010zd}
  G.~Basar, G.~V.~Dunne and D.~E.~Kharzeev,
  Phys.\ Rev.\ Lett.\  {\bf 104} (2010) 232301
  [arXiv:1003.3464 [hep-ph]].

\bibitem{Muller:2010jd}
  B.~M\"uller and A.~Sch\"afer,
  Phys.\ Rev.\  C {\bf 82} (2010) 057902
  [arXiv:1009.1053 [hep-ph]].

\bibitem{Bzdak:2010fd}
  A.~Bzdak, V.~Koch and J.~Liao,
  Phys.\ Rev.\  C {\bf 83} (2011) 014905  
  [arXiv:1008.4919 [nucl-th]].

\bibitem{Wang:2009kd}
  F.~Wang,
  Phys.\ Rev.\  C {\bf 81} (2010) 064902
  [arXiv:0911.1482 [nucl-ex]].

\bibitem{Schlichting:2010qia}
  S.~Schlichting and S.~Pratt,
  Phys.\ Rev.\  C {\bf 83} (2011) 014913  
  [arXiv:1009.4283 [nucl-th]].

\bibitem{Zhang:1999bd}
  B.~Zhang, C.~M.~Ko, B.~A.~Li and Z.~W.~Lin,
  Phys.\ Rev.\  C {\bf 61} (2000) 067901
  [arXiv:nucl-th/9907017].

\bibitem{Lin:2001yd}
  Z.~W.~Lin, S.~Pal, C.~M.~Ko, B.~A.~Li and B.~Zhang,
  Nucl.\ Phys.\  A {\bf 698} (2002) 375
  [arXiv:nucl-th/0105044].

\bibitem{Lin:2004en}
  Z.~W.~Lin, C.~M.~Ko, B.~A.~Li, B.~Zhang and S.~Pal,
  Phys.\ Rev.\  C {\bf 72} (2005) 064901
  [arXiv:nucl-th/0411110].

\bibitem{Wang:1991hta}
  X.~N.~Wang and M.~Gyulassy,
  Phys.\ Rev.\  D {\bf 44} (1991) 3501.

\bibitem{Gyulassy:1994ew}
  M.~Gyulassy and X.~N.~Wang,
  Comput.\ Phys.\ Commun.\  {\bf 83} (1994) 307
  [arXiv:nucl-th/9502021].

\bibitem{Zhang:1997ej}
  B.~Zhang,
  Comput.\ Phys.\ Commun.\  {\bf 109} (1998) 193
  [arXiv:nucl-th/9709009].

\bibitem{Sjostrand:1993yb}
  T.~Sj\"ostrand,
  Comput.\ Phys.\ Commun.\  {\bf 82} (1994) 74.

\bibitem{Li:1995pra}
  B.~A.~Li and C.~M.~Ko,
  Phys.\ Rev.\  C {\bf 52} (1995) 2037
  [arXiv:nucl-th/9505016].

\bibitem{Lin:2001zk}
  Z.~W.~Lin and C.~M.~Ko,
  Phys.\ Rev.\  C {\bf 65} (2002) 034904
  [arXiv:nucl-th/0108039].

\bibitem{Chen:2006vc}
  L.~W.~Chen and C.~M.~Ko,
  Phys.\ Rev.\  C {\bf 73} (2006) 044903
  [arXiv:nucl-th/0602025].

\bibitem{Chen:2006ub}
  J.~H.~Chen {\it et al.},
  Phys.\ Rev.\  C {\bf 74} (2006) 064902.

\bibitem{Zhang:2005ni}
  B.~Zhang, L.~W.~Chen and C.~M.~Ko,
  Phys.\ Rev.\  C {\bf 72} (2005) 024906
  [arXiv:nucl-th/0502056].

\bibitem{Zhang:2000nc}
  B.~Zhang, C.~M.~Ko, B.~A.~Li, Z.~W.~Lin and B.~H.~Sa,
  Phys.\ Rev.\  C {\bf 62} (2000) 054905
  [arXiv:nucl-th/0007003].

\bibitem{Zhang:2002ug}
  B.~Zhang, C.~M.~Ko, B.~A.~Li, Z.~W.~Lin and S.~Pal,
  Phys.\ Rev.\  C {\bf 65} (2002) 054909
  [arXiv:nucl-th/0201038].

\bibitem{Zhang:2006yf}
  B.~Zhang,
  Phys.\ Lett.\  B {\bf 647} (2007) 249
  [arXiv:nucl-th/0606039].

\bibitem{Lin:2002gc}
  Z.~W.~Lin, C.~M.~Ko and S.~Pal,
  Phys.\ Rev.\ Lett.\  {\bf 89} (2002) 152301
  [arXiv:nucl-th/0204054].

\bibitem{Zhang:2007qx}
  S.~Zhang {\it et al.},
  Phys.\ Rev.\  C {\bf 76} (2007) 014904
  [arXiv:0706.3820 [nucl-th]].

\bibitem{Pal:2009zz}
  S.~Pal,
  Phys.\ Rev.\  C {\bf 80} (2009) 041901.

\bibitem{Xu:2010du}
  J.~Xu and C.~M.~Ko,
  Phys.\ Rev.\  C {\bf 83} (2011) 021903.  
 [arXiv:1011.3750 [nucl-th]].

\bibitem{Ma:2010dv}
  G.~L.~Ma and X.~N.~Wang,
  arXiv:1011.5249 [nucl-th].

\bibitem{Voloshin:2004vk}
  S.~A.~Voloshin,
  Phys.\ Rev.\  C {\bf 70} (2004) 057901
  [arXiv:hep-ph/0406311].

\bibitem{Pratt:2010gy}
S.~Pratt,
arXiv:1002.1758 [nucl-th].

\bibitem{Pratt:2010zn}
  S.~Pratt, S.~Schlichting and S.~Gavin,
  arXiv:1011.6053 [nucl-th].

\bibitem{Toneev:2010xt}
  V.~D.~Toneev and V.~Voronyuk,
  arXiv:1012.1508 [nucl-th].


\end{thebibliography}
\end{document}